\documentclass[twocolumn,amsmath,aps]{revtex4}
\usepackage{graphicx}

\newcommand{\be}{\begin{equation}}
\newcommand{\ee}{\end{equation}}
\newcommand{\bea}{\begin{eqnarray}}
\newcommand{\eea}{\end{eqnarray}}

\newcommand{\Eq}[1]{Eq.\,(\ref{#1})}

\newcommand{\ra}{\rangle}
\newcommand{\dg}{\dagger}

\newcommand{\ti}{\tilde}
\newcommand{\mb}{\mbox}

\begin{document}
\draft

\title{ Continuous weak measurement and feedback control of a solid-state charge qubit:
physical unravelling of non-Lindblad master equation }

\author{Shi-Kuan Wang,
Jinshuang Jin \footnote{Present address: Department of Chemistry,
Hong Kong University of Science and Technology, Kowloon, Hong
Kong} and Xin-Qi Li\footnote{E-mail: xqli@red.semi.ac.cn}}
\address{State Key Laboratory for Superlattices and Microstructures,
         Institute of Semiconductors,
         Chinese Academy of Sciences, P.O.~Box 912, Beijing 100083, China }

\date{\today}

\begin{abstract}
Conventional quantum trajectory theory developed in quantum optics
is largely based on the physical unravelling of Lindbald-type
master equation, which constitutes the theoretical basis of
continuous quantum measurement and feedback control. In this work,
in the context of continuous quantum measurement and feedback
control of a solid-state charge qubit, we present a {\it physical}
unravelling scheme of {\it non-Lindblad type} master equation.
Self-consistency and numerical efficiency are well demonstrated.
In particular, the control effect is manifested in the detector
noise spectrum, and the effect of measurement voltage is
discussed. 
\end{abstract}
\maketitle


\section{Introduction}


Quantum trajectory theory has been developed and intensively
applied in quantum optics
\cite{Gisin84,Zoller87,Gisin92,Dalibard92,Carmichael93,Wiseman93,Plenio98}.
However, the theory is largely based on unravelling of
Lindblad-type master equation, which has clear physical
interpretation \cite{Wiseman93}.
For non-Lindblad type master equation \cite{Bre99}, or even the
non-Markovian dissipative dynamics
\cite{Ima94,Str99,Stoc02,Sha04}, unravelling in terms of
stochastic differential equation has also been established for
various purposes. Nevertheless, these schemes lack physical
interpretation, having thus only mathematical meaning.


In solid-state system, the quantum measurement
\cite{Gur97,Schon98,Korotkov99,Averin01,Goan01} of solid-state
qubit and feedback control \cite{Korotkov021,Hop03,Rus95} have
been an extensively studied subject in recent years, being largely
stimulated by the prospect of solid-state quantum computing.
The theoretical description of this solid-state qubit measurement
problem was developed originally in terms of the ``$n$"-resolved
master equation \cite{Gur97,Schon98}, where ``$n$" stands for the
number of electrons passed through the measurement apparatus. 
Alternatively, Bayesian formalism \cite{Korotkov99} and also the
conventional quantum trajectory theory \cite{Goan01} were
developed for this solid-state measurement setup. Being of
interest, it can be shown that all these three approaches are
precisely equivalent to each other .

More specifically, for the setup of a pair of coupled quantum dots
(CQDs) qubit measured by a quantum point contact (QPC) detector,
which has become an experimentally studied system \cite{Hay0304},
we found that the above mentioned theories were restricted to the
limit of large measurement voltage across the QPC
\cite{Li04,Li05}. 
At finite voltages, i.e., as the measurement voltage is comparable
to or not much higher than the qubit's intrinsic energy scale, the
measurement dynamics is governed by a non-Lindblad type master
equation \cite{Li04}. In this regime, the ``$n$"-resolved master
equation was also developed to study the readout characteristics
\cite{Li05}. 
In the present work, we further extend it to a quantum trajectory
theory, which conditions the state evolution on the {\it entire
measurement records}. It is well known that this kind of
description is essential to the quantum feedback control and other
possible applications.

The paper is organized as follows.
In Sec.\ II we first present a brief description for the setup
under study, then outline the ``n"-resolved master equation for
the measurement. 
In connection with the conditional state evolution, physical
unravelling scheme is constructed and practical Monte Carlo
simulation is carried out. 
Also, for completeness and latter use, a generalized quantum-jump
approach is formulated for the calculation of noise spectrum. 
For the sake of brevity, some complicated expressions and
mathematical details are put in two Appendices.
In Sec.\ III we apply the developed formalism to the study of
feedback control, based on a suboptimal feedback algorithm.
Finally, conclusion is presented in Sec.\ IV.

\section{ Continuous Measurement of a Charge Qubit}

\subsection{Model Description}

Let us consider a solid-state charge qubit measured by a nearby
mesoscopic detector.
In this work, as well studied in literature
\cite{Gur97,Korotkov99,Averin01,Goan01},
the charge qubit is modelled by
a pair of coherently coupled quantum dots with an extra electron
in it, and the mesoscopic detector can be a mesoscopic quantum
point contact (QPC).
The Hamiltonian of this qubit-plus-detector setup reads
\begin{subequations}
\begin{eqnarray}
H   &=&  H_0+H'
\\
H_0 &=&  H_{qb}   + \sum_{k}(\epsilon^L_k
c^\dagger_kc_k+\epsilon^R_k d^\dagger_kd_k)
\\
H_{qb} &=& \epsilon_a|a\rangle\langle a|+
\epsilon_b|b\rangle\langle b|+\Omega(|b\rangle\langle
a|+|a\rangle\langle b|)
\\
H'  &=& \sum_{k,q} [(\mathcal{T}_{kq}+\chi_{kq}|a\rangle\langle
a|)
         c^{\dg}_k d_q  + \mb{H.c.}] .
\end{eqnarray}
\end{subequations}
In this decomposition, the free part of the total Hamiltonian,
$H_0$, contains Hamiltonians  of the measured qubit $H_{\rm qb}$
and the QPC's reservoirs. The operators $c_k^{\dag}(c_k)$ and
$d_k^{\dag}(d_k)$ are, respectively, the electronic creation
(annihilation) operators of the QPC's left and right reservoirs.
The qubit states $|a\rangle$ and $|b\rangle$ correspond to the
electron locating in the left and right dots.
Introducing $\epsilon=(\epsilon_a-\epsilon_b)/2$ and taking
$(\epsilon_a+\epsilon_b)/2$ as the reference energy, the qubit
eigenenergies read $E_1=\sqrt{\epsilon^2+\Omega^2}=\Delta/2$, and
$E_2=-\sqrt{\epsilon^2+\Omega^2}=-\Delta/2$. Correspondingly, the
eigenstates are
$|1\rangle=\cos\frac{\theta}{2}|a\rangle+\sin\frac{\theta}{2}|b\rangle$
for the excited state, and
$|0\rangle=\sin\frac{\theta}{2}|a\rangle-\cos\frac{\theta}{2}|b\rangle$
for the ground state, where $\theta$ is defined by
$\cos\theta=2\epsilon/\Delta$, and $\sin\theta=2\Omega/\Delta$.

For the sake of simplicity, we assume
$\mathcal{T}_{kq}=\mathcal{T}$ and $\chi_{kq}=\chi$,  i.e., the
tunneling amplitudes are real and reservoir-state independent.
Corresponding to the qubit states $|a\rangle$ and $|b\rangle$, the
stationary detector currents read $I_a=2\pi
g_Lg_R(\mathcal{T}+\chi)^2V$, and $I_b=2\pi g_Lg_R\mathcal{T}^2V$,
respectively, where $g_{L(R)}$ is the density of state of the left
(right) reservoir, and $V$ is the measurement voltage.
Physically, $\Delta I=I_a-I_b$ characterizes the detector's
response to the qubit electron's location in the CQDs. The
detector is said in the so-called weakly responding regime if
$\Delta I\ll I_0=(I_a+I_b)/2$. In this work we assume this regime,
which enables us to ignore the individual electron tunnelling
events and treat the current as a continuous diffusive variable.

\subsection{``$n$"-Resolved Master Equation}

The so-called ``$n$"-resolved master equation is obtained by {\it
partially} tracing out the detector's microscopic degrees of
freedom but keeping track of the number ``$n$" of electrons that
have tunnelled through the detector during the time period $(0,t)$
\cite{Gur97,Schon98,Li05}. Originally, it was derived in Ref.\
\onlinecite{Gur97} from the many-body Schr\"{o}dinger equation at
zero temperature and in the large measurement voltage limit.
Later, it was proved that this approach is completely equivalent
to the Bayessian approach \cite{Korotkov99} and the quantum
trajectory theory \cite{Goan01}, which share the same Lindblad
type master equation and its unravelling.

Alternatively, under arbitrary (i.e. not high enough) voltages, it
was found that this measurement problem {\it cannot} be described
by a Lindblad type master equation \cite{Li04,Li05}. In
particular, a non-Lindblad type master equation was derived
\cite{Li04}, and its ``$n$"-resolved counterpart reads \cite{Li05}
\begin{eqnarray}
\dot{\rho}^{(n)}
&=&-i\mathcal{L}\rho^{(n)}-\frac{1}{2}\{Q\tilde{Q}\rho^{(n)}
-\tilde{Q}^{(-)}\rho^{(n-1)}Q\nonumber\\&
&-\tilde{Q}^{(+)}\rho^{(n+1)}Q+\mb{H.c.}\} . \label{n-ME-1}
\end{eqnarray}
Here, $Q=\mathcal{T}+\chi|a\rangle\langle a|$,
$\ti{Q}=\ti{Q}^{(+)}+\ti{Q}^{(-)}$,
$\ti{Q}^{(\pm)}=\ti{C}^{(\pm)}({\cal L})Q$, and
$\ti{C}^{(\pm)}({\cal L})=\int^{\infty}_{-\infty} dt C^{(\pm)}(t)
e^{-i{\cal L}t}$. $C^{(\pm)}(t)$ are the reservoir electron
correlation functions. Under wide-band approximation for the QPC
reservoirs, the spectral function $\ti{C}^{(\pm)}({\cal L})$ can
be explicitly carried out as $\ti{C}^{(\pm)}({\cal L})
  =  2\pi g_Lg_R \left[x/(1-e^{-x/T}) \right]_{x=-{\cal L}\mp V}$,
where $T$ is the reservoir temperature (in this work we use the
unit system of $\hbar=e=k_B=1$).
It is of interest to note that the Liouvillian operator
$\mathcal{L}$ in $\ti{C}^{(\pm)}({\cal L})$ contains the
information of energy exchange between the detector and the qubit,
which correlates the energy relaxation of the measured qubit with
the inelastic electron tunnelling in the detector.
Note also that in the derivation of the above ``n"-resolved master
equation we did not make assumption of large bias voltage across
the QPC detector. At large voltage limit, i.e., the bias voltage
is much larger than the internal energy scale of the qubit, the
spectral function $\ti{C}^{(\pm)}({\cal L})
\simeq\ti{C}^{(\pm)}({0})$, and \Eq{n-ME-1} reduces to the result
obtained in Ref.\ \onlinecite{Gur97}.

Formally, we rewrite Eq. (\ref{n-ME-1}) as
\begin{equation}
\dot{\rho}^{(n)} =-i\mathcal{L}\rho^{(n)}-\mathcal{R}\rho^{(n)}
+\mathcal{R}_1\rho^{(n-1)}+\mathcal{R}_2\rho^{(n+1)} ,
\label{n-ME-2}
\end{equation}
where $\mathcal{R}$, $\mathcal{R}_1$ and $\mathcal{R}_2$ are
superoperators defined in accord with \Eq{n-ME-1}. 
To solve this infinite number of coupled equations, we perform the
discrete Fourier transformation, $\rho(k,t)=\sum_n
e^{ink}\rho^{(n)}(t)$, yielding
\begin{equation}
\dot{\rho}(k,t) = \left[ -i\mathcal{L}-\mathcal{R}
 +e^{ik}\mathcal{R}_1+e^{-ik}\mathcal{R}_2\right] \rho(k,t) .
 \label{k-ME-1}
\end{equation}
Explicitly, in the localized dot-state representation
$\{|a\rangle,|b\rangle\}$ we obtain
\begin{widetext}
\begin{equation}\label{k-ME-2}
\begin{pmatrix}
\dot{\rho}_{aa}\\\dot{\rho}_{bb}\\\dot{\rho}_{ab}\\\dot{\rho}_{ba}
\end{pmatrix}
=
\begin{pmatrix}
a_1 &0 &a_2+i\Omega
&a_2-i\Omega\\0&b_1&b_2-i\Omega&b_2+i\Omega\\c_3+i\Omega&c_2-i\Omega&c_1
-i\epsilon_a+i\epsilon_b&0\\c_3-i\Omega&c_2+i\Omega&0&c_1+i\epsilon_a-i\epsilon_b
\end{pmatrix}
\begin{pmatrix}
\rho_{aa}\\\rho_{bb}\\\rho_{ab}\\\rho_{ba} .
\end{pmatrix}
\end{equation}
\end{widetext}
For brevity, the explicit expressions of the coefficients
$a_{1(2)}$, $b_{1(2)}$ and $c_{1(2,3)}$ are ignored here and are
put alternatively in Appendix A.

Formally, we reexpress the Fourier-transformed master equation as
$\dot{\rho}(k,t)={\cal M}(k)\rho(k,t)$, and the solution reads
$\rho(k,t)=e^{{\cal M}(k)(t-t_0)}\rho(k,t_0)$. Note that we are
concerned with the ``n"-resolved state evolution from $t_0$ to
$t$, i.e., the counting of ``n" starts from the moment $t_0$. We
thus have $\rho^{(n)}(t_{0})=\rho(t_{0})\delta_{n,0}$, and
$\rho(k,t_0)=\rho(t_{0})$.
With the knowledge of $\rho(k,t)$, the inverse Fourier transform
gives \bea\label{rho-n-t} \rho^{(n)}(t)=\sum_{k}e^{-ink}\rho(k,t)
    =\sum_{k}e^{-ink}e^{{\cal M}(k)(t-t_0)}\rho(t_0) .
\eea Strikingly, we can introduce a propagator for the state
evolution, ${\cal U}(n,t)=\sum_{k}e^{-ink}e^{{\cal M}(k)t}$. Since
this propagator is completely determined by the {\it dynamic
structure} of the master equation but does not depend on the
initial state, we can numerically evaluate it by a ``one-time
task" via such as the fast Fourier transformation. 
This feature leads to a very efficient Monte Carlo simulation for
the measurement-history conditioned evolution (i.e. the quantum
trajectory simulation).

\subsection{Monte Carlo Simulation for Conditional Evolution}

It seems that $\rho^{(n)}(t)$ contains {\it less} information
about the measurement record (history) than the conditional state
$\rho_c(t)$ in the conventional quantum trajectory theory
\cite{Goan01}, since it only implies that {\it totally} there have
been $n$ electrons passed through the QPC junction during the
specified time period. 
However, if we make successive readout for the electron numbers
``$n_k$" passed through the detector during the time interval
$(t_{k-1},t_k)$, we actually record the measurement current
$I_c(t)$ of a single realization. After each time of reading out
``$n_k$", the statistically mixed state $\rho^{(n)}(t_{k})$ with
any possible ``$n$" would ``collapse" to a normalized state
$\rho^{(n_k)}(t_{k})$ with definite ``$n=n_k$". The set of records
$\{n_k: k=1,2,\cdots\}$ corresponds to the current $I_c(t)$, and
the set of states $\{\rho^{(n_k)}(t_{k}): k=1,2,\cdots\}$ is
nothing but the conditional state $\rho_c(t)$ in the quantum
trajectory theory.

This {\it state-update} procedure based on the ``$n$"-resolved
master equation was introduced in Ref.\ \onlinecite{Korotkov99},
where its exact equivalence to the Bayessian and quantum
trajectory theories was analytically proved. That is, the {\it
conditional} master equation can be re-derived based on the
``$n$"-resolved master equation together with the above
``collapse" idea.
However, we found that this can be done only in the large voltage
limit which leads to a Lindblad-type master equation
\cite{Korotkov99,Goan01}. For arbitrary voltage, rather than
deriving a conditional master equation to describe the
measurement-record conditioned evolution, we would like here to
develop an efficient numerical unravalling scheme which has the
advantage of being applicable to non-Lindblad type master equation
as studied in this work.

More quantitatively, let us consider the state evolution during
$[t_j,t_j+\tau]$. That is, starting with a definite state at
$t_j$, say $\rho(t_j)$, the state $\rho^{(n_j)}(t_j+\tau)$ at
$t_j+\tau$ can be calculated via
\begin{equation} \label{rho-c-tj}
\rho^{(n_j)}(t_j+\tau)={\cal U}(n_j,\tau)\rho(t_j) .
\end{equation}
If the measurement is made but the result is ignored, the
(mixture) state is described by
\begin{eqnarray}\label{rho-tj}
\rho(t_j+\tau)=\sum_{n_j}\rho^{(n_j)}(t_j+\tau)
  = \sum_{n_j} {\rm Pr}(n_j)\rho_c(n_j,t_j+\tau) ,
\end{eqnarray}
where ${\rm Pr}(n_j)={\rm Tr}[\rho^{(n_j)}(t_j+\tau)]$ stands for
the probability having $n_j$ electrons tunnelled through the
detector, and $\rho_c(n_j,t_j+\tau)=\rho^{(n_j)}(t_j+\tau)/{\rm
Pr}(n_j)$ is the normalized state conditioned by the definite
number of $n_j$ electrons observed passed through the detector.

The second equality of \Eq{rho-tj} implies that if we
stochastically generate $n_j$ according to the probability ${\rm
Pr}(n_j)$ for each time interval $[t_j,t_j+\tau]$, step by step
from $t_0$ to $t$, and ``collapse" the state definitely onto
$\rho_c(n_j,t_j+\tau)$, i.e.,
$\rho_c(t_j+\tau)=\rho^{(n_j)}(t_j+\tau)/{\rm Pr}(n_j) ={\cal
U}(n_j,\tau)\rho(t_j)/{\rm Tr}[{\cal U}(n_j,\tau)\rho(t_j)]$, we
have in fact simulated a particular realization for the selective
state evolution conditioned on the (continuous) specific
measurement result. 
The simple ensemble average over a large number of particular
realizations of $\rho_c(t)$ recovers the un-conditional state
$\rho(t)$. 
Obviously, this unravelling scheme is completely equivalent to the
spirit of the conventional quantum trajectory theory, despite that
in this context we are unable to derive an explicit stochastic
differential equation to unravel the underlying non-Lindblad
master equation. However, they have precisely the same physical
meaning.

To stochastically generate $n_j$ according to the probability
${\rm Pr}(n_j)$, two procedures are adopted as follows: 
(i) Based on the ensemble average current $I(t)={\rm Re}\{{\rm
Tr}[\bar{Q}\rho(t) Q]\}$, which was derived in Ref.
\onlinecite{Li05}, the output current in particular measurement
realization reads
\begin{equation}\label{Ict}
I_c(t)={\rm Re}\{{\rm Tr}[\bar{Q}\rho_c(t) Q]\}+\xi(t).
\end{equation}
The first term in this equation is related to the qubit dynamics
of {\it conditional evolution}. 
The second noisy term $\xi(t)$ originates from the detector's {\it
intrinsic} noise, which is a Poisonian variable in the regime of
point process, and a Gaussian variable in the diffusive regime. In
the latter case, $\xi(t)$ has zero mean value, and the spectral
density $S_\xi=2I_0\coth\frac{V}{2T}$. 
At zero temperature and large voltage limit, this treatment
recovers the existing result of quantum trajectory theory and
Bayesian approach \cite{Korotkov99,Goan01}, i.e.,
$I_{c}(t)=\rho_{c,aa}(t)I_a+\rho_{c,bb}(t)I_b+\xi(t)$, with
$S_\xi=2I_0$.
(ii) Straightforwardly, in our simulation we relate the {\it
stochastic} electron number $n_j$ with $I_{c}(t)$ via
$n_j=\int^{t_j+\tau}_{t_j}dt'I_c(t')=\bar{I}_c(t_j)\tau+dW(t_j)$,
where $\bar{I}_c(t_j)={\rm Re}\{{\rm Tr}[\bar{Q}\rho_c(t_j) Q]\}$,
and $dW(t_j)$ is the Wiener increment during $[t_j,t_j+\tau]$.


In Fig.\ 1 we plot a comparison of the ensemble average of the
Monte Carlo simulation (over 500 quantum trajectories) with the
result directly given by the unconditional master equation. The
excellent agreement shows the validity and efficiency of the
proposed unravelling scheme.
\begin{figure}
\begin{center}
\includegraphics*[width=8cm,height=8cm,keepaspectratio]{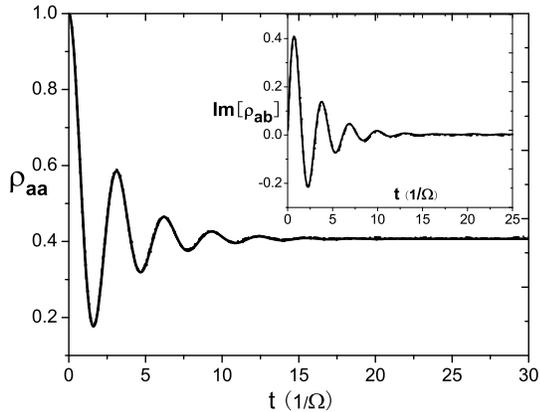}
\caption{\label{comparison}
Ensemble average of the Monte Carlo simulation over 500 quantum trajectories
(dash-dotted line) versus the result directly given by the unconditional
master equation (solid line).
It is assumed that the initial state of the qubit $|\psi\ra=|a\ra$.
The relevant parameters are:
$\mathcal{T}=20 \Omega$, $\chi=0.7\Omega$, $V=0.5\Omega$,
$\epsilon=0.25\Omega$, $T=1.0\Omega$,
and $g_{L(R)}=1/\sqrt{2\pi}\Omega$.}
\end{center}
\end{figure}
In this context, two points are likely to be highlighted: (i) the
measurement voltage considered here is moderately finite, but not
the high voltage limit \cite{Korotkov99,Goan01}; (ii) the
corresponding non-Lindblad master equation is unravelled {\it
physically}, having the same physical interpretation as provided
by Wiseman {\it et al} \cite{Wiseman93}.

In Fig.\ 2 we show the main features of the conditional state
evolution.
(i) Assuming that the coherent coupling is switched off
($\Omega=0$), in Fig.\ 2(a) we illustrate the wave-function
collapse of a pure state $|\psi\ra=1/\sqrt{2}(|a\ra+|b\ra)$ under
measurement. 
This feature has deep implication in understanding the {\it
measurement postulate} in quantum mechanics, which has been
highlighted by the concept of {\it gradual collapse} for a typical
solid-state two-level state under (weak) measurement, as discussed
in particular by Leggett \cite{Leggett83}. 
The reason for this {\it gradual collapse} is the weak coupling
and the finite noise of the detector, which make the quantum
measurement need some time until acceptable signal-to-noise ratio
is reached.
(ii) On the other hand, if $\Omega\neq0$, the ideal measurement
will lead to the {\it gradual purification} of the qubit state
starting, for instance, with a completely mixed state. This is
shown by the revival of the coherent Rabi oscillation in Fig.\
2(b).
(iii) As shown in Fig.\ 3(c), with the increase of the measurement
strength ($|\chi|/\Omega$), the duration time on each qubit state
is enhanced, while the switching time between them is reduced.
This is an obvious signature of the quantum Zeno effect, appearing
in the regime of {\it gradual} but not the conventional {\it
instantaneous} collapse.
(iv) In the conditional dynamics, the output current $I_{c}(t)$
can basically follow the conditional qubit state, as shown by
Fig.\ 2(d). Due to the intrinsic noise of the detector, here the
filtered current, i.e., $\bar{I}_c(t)=\frac{1}{\Delta
t}\int^{t+\Delta t}_{t} I_{c}(t^{'})dt^{'}$, is plotted, where
$\Delta t$ is the ``filtering window".
\begin{figure}
\begin{center}
\includegraphics*[width=8cm,height=8cm,keepaspectratio]{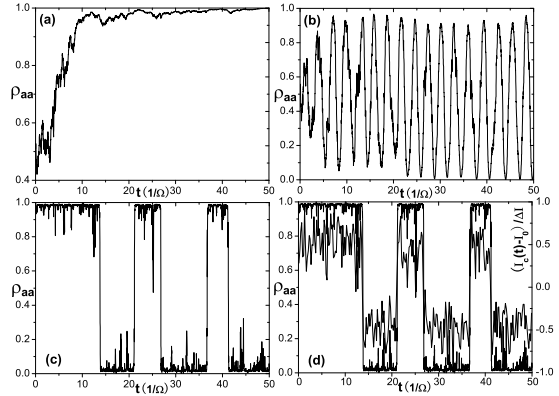}
\caption{\label{trajectory}
The main characteristics of the conditional state evolution
under continuous measurement:
(a) gradual localization ($\Omega=0$) from an initial
superposition state $|\psi\ra=1/\sqrt{2}(|a\ra+|b\ra)$;
(b) gradual purification from a completely mixed state;
(c) zeno effect under relatively strong continuous measurement;
and (d) conditional state evolution (thick line) versus
the filtered output current (thin line), which is obtained by
$\frac{1}{\Delta t}\int^{t+\Delta t}_{t} I_{c}(t^{'})dt^{'}$,
with the filtering window $\Delta t=0.2/\Omega$.
The parameters ($\mathcal{T}$, $\chi$, $V$, $\epsilon$, $T$) are
adopted as (in unit of $\Omega$):
(a)(20.0, 0.13, 3.0, 0.25, 1.0),
(b)(20.0, 0.7, 3.0, 0.25, 1.0), (c)(25.0, 4.0, 3.0, 0.25, 1.0),
and (d)(25.0, 4.0, 3.0, 0.25, 1.0).
The density of states $g_{L(R)}=1/\sqrt{2\pi}\Omega$.    }
\end{center}
\end{figure}


\subsection{Noise Spectrum }

In the same spirit of conventional quantum trajectory theory
\cite{Wiseman93,Goan01}, the present unravelling scheme also
provides a natural way to calculate the output power spectrum. The
details of derivation is referred to Appendix B, here we simply
present the resultant expression of the output current correlator,
which reads
\begin{eqnarray}
K_{I}(\tau)&\equiv&E[I_{c}(t+\tau)I_{c}(t)]-E[I_{c}(t+\tau)]E[I_{c}(t)]
\nonumber
\\
&=&{\rm
Tr}[\overline{\mathcal{U}}e^{\mathcal{L}\tau}\overline{\mathcal{U}}\rho(t)]
-{\rm Tr}[\overline{\mathcal{U}}\rho(t+\tau)]{\rm
Tr}[\overline{\mathcal{U}}\rho(t)]\nonumber
\\
& &+{\rm Tr}[\overline{\mathcal{U'}}\rho(t)]\delta(\tau),
 \end{eqnarray}
where $\overline{\mathcal{U}}=\sum_{n}n\mathcal{U}(n,dt)/dt$,
$\overline{\mathcal{U'}}=\sum_{n}n^{2}\mathcal{U}(n,dt)/dt$,
$\rho(t+\tau)=e^{\mathcal{L}\tau}\rho(t)$. For stationary state£¬
$K_{I}(\tau)={\rm Tr}[\overline{\mathcal{U}}e^{\mathcal{L}\tau}
\overline{\mathcal{U}}\rho(\infty)]-{\rm
Tr}[\overline{\mathcal{U}} \rho(\infty)]^{2}+{\rm
Tr}[\overline{\mathcal{U'}}\rho(\infty)]\delta(\tau)$.

In practice, particularly in the presence of quantum feedback, the
ensemble averaged evolution represented by $e^{\mathcal{L}\tau}$
can be implemented by numerically averaging the stochastic
trajectories. 
In the absence of quantum feedback, the above quantum trajectory
approach can precisely recover the analytic result of stationary
state noise spectrum, obtained by using the MacDonald formula
\cite{Li05}.

\section{Quantum Feedback Control}

Quantum feedback control is one of the typical means of quantum
coherence control. In quantum optics, the study of quantum
feedback control has been going on for more than a decade
\cite{Wiseman1993}. However, it is a relatively new subject in
solid states \cite{Korotkov021,Hop03,Rus95}.
In particular, the conditional state evolution under continuous
weak measurement has been {\it experimentally} demonstrated in
solid-state qubit very recently \cite{Kat06}. This may pave a way
to the quantum feedback control in solid-states.
For the solid-state setup under present study, we now consider the
feedback control of the qubit coherent evolution, by unravelling
the underlying measurement dynamics that is in general governed by
non-Lindblad master equation.

The basic idea is to convert the measurement information in the
output current $I_{c}(t)$ into the evolution of qubit state
$\rho_{c}(t)$. By comparing $\rho_{c}(t)$ with the desired state
$\rho_{d}(t)$, their difference is then employed to modify the
qubit Hamiltonian in order to reduce their difference in next step
evolution.
Specifically, we consider a symmetric qubit (i.e. $\epsilon=0$).
The desired state is $|\psi_{d}(t)\rangle=\cos(\Omega
t)|a\rangle-i\sin(\Omega t)|b\rangle$. In real-time feedback
control, each successive feedback acts only for an infinitesimal
time interval $\triangle t$. In the so-called Bayessian
state-estimate-based feedback, the suboptimal algorithm is
desirable \cite{Doherty01,Li06}. That is, the algorithm is
constructed such that the state evolution in each infinitesimal
time step will maximize the fidelity of the estimated state with
the desired (target) state.
In more detail, as far as the term related to the feedback
Hamiltonian is concerned, the state $\rho_{c}(t+\triangle t)$ is
given by
\begin{eqnarray}
 \rho_{c}(t+\triangle t)&=&
 \rho_{c}(t)-i[H_{fb},\rho_{c}(t)]\triangle
 t\nonumber
 \\
 & &-\frac{1}{2}[H_{fb},[H_{fb},\rho_{c}(t)]](\triangle t)^{2}+\cdots ~.
\end{eqnarray}
The fidelity of this state with the target state reads

\begin{multline}
\langle \psi_{d}(t) | \rho_{c}(t+\triangle t) |\psi_{d}(t) \rangle
 \\
\shoveleft{ = \langle \psi_{d}(t) | \rho_{c}(t) |\psi_{d}(t)
\rangle -i \langle \psi_{d}(t) | [H_{fb},\rho_{c}(t)] |\psi_{d}(t)
\rangle \triangle t}
\\- \frac{1}{2}\langle \psi_{d}(t) | [H_{fb},[H_{fb},\rho_{c}(t)]]
|\psi_{d}(t) \rangle (\triangle t)^{2}+\cdots ~.
\end{multline}
To optimize the fidelity, one should maximize the coefficient of
$\triangle t$, which is the dominant term. Similar to other
control theories, the maximization must be subject to certain
constraints, e.g., the restriction on the maximum eigenvalue of
$H_{fb}$, the sum of the norms of
 the eigenvalues, or the sum of the squares of the eigenvalues, etc.
 Physically, these constraints stem from the
 limitation of the feedback strength or finite Hamiltonian
 resources. Here we adopt the last type of constraint,
 namely,
${\rm Tr}[H_{fb}^{2}]\leq \mu ~$. Under this constraint, the
feedback Hamiltonian can be constructed in terms of
\begin{eqnarray}
 H_{fb}=i\lambda[|\psi_{d}(t)\rangle\langle\psi_{d}(t)|,\rho_{c}(t)] ~,
\end{eqnarray}
 where $\lambda=\sqrt{\frac{\mu}{2(a-b^{2})}}$,
 with $a =\langle\psi_{d}(t)|\rho_{c}(t)^{2}|\psi_{d}(t)\rangle$,
 and $b=\langle\psi_{d}(t)|\rho_{c}(t)|\psi_{d}(t)\rangle$.

Combining the above feedback Hamiltonian with the previously
developed state unravelling scheme, the estimated state
$\rho_{c}(t)$ can be straightforwardly calculated, leading to the
state propagation in the presence of feedback.
Figure 3 shows the control result with quantum feedback, where the
ensemble-average has been made over large number of Monte-Carlo
simulated trajectories. Here the measurement voltage is quite
moderate, i.e., $V=3 \Omega$, which is beyond the theoretical
description in large-voltage limit as previously studied
\cite{Gur97,Korotkov99,Goan01,Korotkov021,Hop03,Rus95}. We observe
that the control effect is evident: by increasing the feedback
strength $\lambda$ the measurement induced back-action can be
largely eliminated, and the desired coherent oscillation of the
qubit can be maintained {\it for arbitrarily long time}.
\begin{figure}
\begin{center}
\includegraphics*[width=8cm,height=7cm,keepaspectratio]{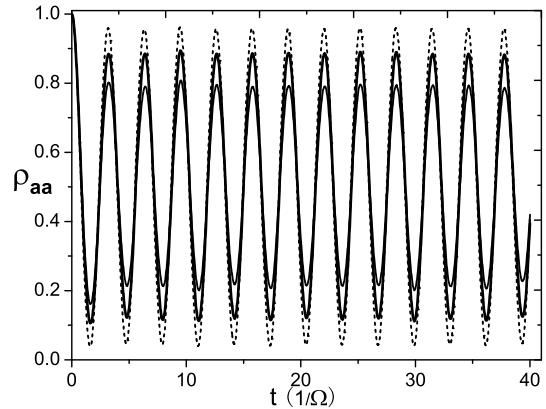}
\caption{\label{feedback}
Feedback control of the coherent oscillation of qubit state,
resulting from feedback strengths $\lambda=0.5$ (thin solid line),
$1.0$ (thick solid line), and $3.5$ ( dashed line).
The parameters are:
$\mathcal{T}=20.0 \Omega$, $\chi=0.7\Omega$, $V=3.0\Omega$,
$\epsilon=0.0\Omega$, $T=1.0\Omega$,
and $g_{L(R)}=1/\sqrt{2\pi}\Omega$. }
\end{center}
\end{figure}



It will be of interest to compare the quantum measurement in the
presence of feedback to the well-known quantum non-demolition
(QND) measurement. For the solid-state qubit, elegant schemes of
QND measurement have been proposed very recently
\cite{Averin02,Butticker05}.
Here, for the qubit measurement under consideration, we
demonstrate that the QND measurement is equivalent to the usual
back-action-present measurement plus quantum feedback. In Fig.\ 4
we present the calculated output power spectrum of the QPC
detector. The peak at $\omega=\Omega$ indicates the coherent
oscillation of the qubit. In the absence of feedback, it has been
shown that the peak-to-background ratio cannot be larger than 4
\cite{Averin01}, due to the back-action of measurement. In the
presence of feedback, however, we obtain very sharp peak here
which indicates almost ideal coherent oscillations.
\begin{figure}
\begin{center}
\includegraphics*[width=8cm,height=7cm,keepaspectratio]{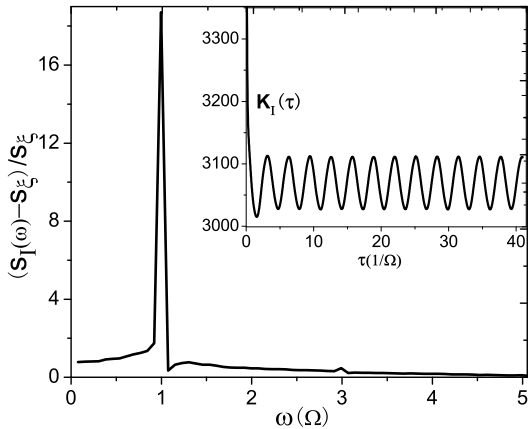}
\caption{\label{feedbacknoise}
Narrowed coherence peak in the noise spectrum
as an indicator for the feedback effect
(Inset: the coherently oscillating correlation function).
The parameters are : $\mathcal{T}=10.0 \Omega$,
$\chi=0.2\Omega$, $V=3.0\Omega$, $\epsilon=0.0\Omega$,
$T=0.5\Omega$, $\lambda=15$,
and $g_{L(R)}=1/\sqrt{2\pi}\Omega$.}
\end{center}
\end{figure}
Theoretically, since no steady-state is available in the presence
of feedback, the start time of the qubit evolution is chosen as
the {\it initial time} of the current correlation function, and
the noise spectrum is the Fourier transform of the correlate
function with respect to the later time (difference). 
Experimentally, this feedback-induced sharper peak can be employed
as an indicator for the feedback effect in practice. We expect
that this kind of experiment can be performed in the not-far
future.

Finally, we address the effect of measurement voltage. Figure 5
shows the synchronization degree of the feedback versus the
measurement voltage. 
It is of interest to notice that there exists an {\it optimal}
measurement voltage for relatively small feedback strength. This
is because for larger voltage the back-action is relatively too
strong, while for smaller voltage the information of the measured
state cannot be extracted out efficiently. As a result, the
turnover behavior of the synchronization degree versus voltage is
found. 
However, with the increase of the feedback strength, the strong
back-action can be eliminated more efficiently by the feedback. In
this case, the synchronization degree does not decrease
considerably as the measurement voltage increases, as shown in
Fig.\ 5.
\begin{figure}
\begin{center}
\includegraphics*[width=8cm,height=7cm,keepaspectratio]{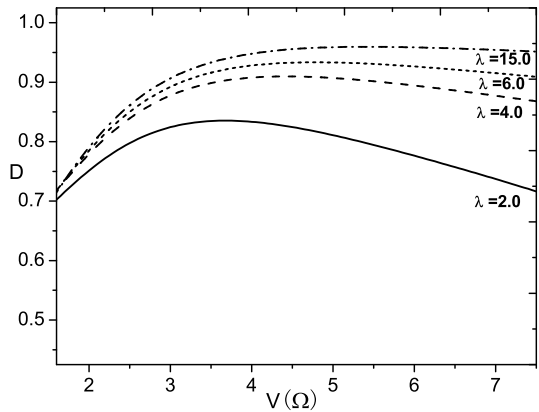}
\caption{\label{DandVoltage} Effect of measurement voltage on the
feedback control. Here the quantity {\it synchronization degree},
which is defined as $D=2\langle {\rm
Tr}[\rho_{c}(t)\rho_{d}(t)]\rangle-1$, with $\langle \cdots
\rangle$ meaning the average over time, is employed to
characterize the control quality. Note that for perfect control
the synchronization degree is unity (i.e., $D=1$). The parameters
are: $\mathcal{T}=20.0 \Omega$, $\chi=0.7\Omega$,
$\epsilon=0.0\Omega$, $T=1.0\Omega$, and
$g_{L(R)}=1/\sqrt{2\pi}\Omega$. }
\end{center}
\end{figure}

\section{Concluding Remarks}

To summarize, in the context of continuous quantum measurement and
coherence control of solid-state charge qubit, we have presented a
unravelling scheme for the non-Lindblad type master equation.
Based on it, we also constructed an efficient method to calculate
the noise spectrum,  which can be regarded as a generalization of
the standard quantum jump theory developed in quantum optics. 
Despite the absence of analytic formalism, the numerical
implementation in practice was demonstrated to be straightforward
and efficient. Illustrative application was further contributed to
the quantum feedback control under arbitrary measurement voltages.
The detector noise spectrum under feedback was calculated, and its
narrowing clearly reflected the control effect. Also, the effect
of measurement voltage was discussed.

The present study has been focused on the setup of double-dot
qubit measured by quantum point contact. However, for other
solid-state setup such as charge qubit measured by
single-electron-transistor, similar unravelling scheme can be
constructed. Owing to the fact that in general (e.g. in the
presence of many-body Coulomb correlations) the measurement
dynamics is not governed by Lindblad type master equation, the
present ``n"-resolved master equation based unravelling scheme
seems quite desirable.
Finally, in spite of the various unravelling schemes for
non-Lindblad type master equation or even for non-Markovian
dissipative systems, to our knowledge all of them are largely {\it
mathematical}. Therefore, the present {\it physical} unravelling
scheme is of interest and valuable, which may find
applications in the field of solid-state quantum information.

\vspace{3ex} {\it Acknowledgments.} Support from the National
Natural Science Foundation of China (No.\ 90203014, 60376037, and
60425412), the Major State Basic Research Project No.\ G001CB3095
of China, and the Research Grants Council of the Hong Kong
Government is gratefully acknowledged.
\appendix
\section{matrix elements of $\mathcal{M}(k)$}
The coefficients in Eq.(5) read
\begin{subequations}
\begin{eqnarray}
a_1&=&-Q_{aa}[\tilde{Q}_{aa}-e^{ik}\tilde{Q}^{(-)}_{aa}-e^{-ik}\tilde{Q}^{(+)}_{aa}],
\\
a_2&=&-\frac{Q_{aa}}{2}[\tilde{Q}_{ab}-e^{ik}\tilde{Q}^{(-)}_{ab}-e^{-ik}\tilde{Q}^{(+)}_{ab}],
\\
b_1&=&-Q_{bb}[\tilde{Q}_{bb}-e^{ik}\tilde{Q}^{(-)}_{bb}-e^{-ik}\tilde{Q}^{(+)}_{bb}],
\\
b_2&=&-\frac{Q_{bb}}{2}[\tilde{Q}_{ba}-e^{ik}\tilde{Q}^{(-)}_{ba}-e^{-ik}\tilde{Q}^{(+)}_{ba}],
\\
c_1&=&-\frac{1}{2}[(Q_{aa}\tilde{Q}_{aa}+Q_{bb}\tilde{Q}_{bb})
-e^{ik}(Q_{aa}\tilde{Q}^{(-)}_{bb}+Q_{bb}\tilde{Q}^{(-)}_{aa})\nonumber
\\
&&-e^{-ik}(Q_{aa}\tilde{Q}^{(+)}_{bb}+Q_{bb}\tilde{Q}^{(+)}_{aa})],
\\
c_2&=&-\frac{1}{2}[Q_{aa}\tilde{Q}_{ab}-e^{ik}Q_{bb}
\tilde{Q}^{(-)}_{ab}-e^{-ik}Q_{bb}\tilde{Q}^{(+)}_{ab}] ,
\\
c_3&=&-\frac{1}{2}[Q_{bb}\tilde{Q}_{ba}-e^{ik}Q_{aa}
\tilde{Q}^{(-)}_{ba}-e^{-ik}Q_{aa}\tilde{Q}^{(+)}_{ba}] ,
\end{eqnarray}
\end{subequations}
where
\begin{subequations}
\begin{eqnarray}
\tilde{Q}^{(\pm)}_{aa}&=&\tilde{C}^{(\pm)}(0)[\textit{T}
+\frac{1}{2}\chi(1+\cos^2\theta)]+\chi\lambda_{\pm}\sin^2\theta,\nonumber
\\
\tilde{Q}^{(\pm)}_{bb}&=&\tilde{C}^{(\pm)}(0)[\textit{T}
+\frac{1}{2}\chi(1+\sin^2\theta)]-\chi\lambda_{\pm}\sin^2\theta
,\nonumber
\\
\tilde{Q}^{(\pm)}_{ab}&=&\frac{1}{2}\chi\tilde{C}^{(\pm)}(0)
\sin\theta\cos\theta+\chi\sin\theta(\bar{\lambda}_{\pm}-\lambda_{\pm}\cos\theta),\nonumber
\\
\tilde{Q}^{(\pm)}_{ba}&=&\frac{1}{2}\chi\tilde{C}^{(\pm)}(0)
\sin\theta\cos\theta-\chi\sin\theta(\bar{\lambda}_{\pm}+\lambda_{\pm}\cos\theta),\nonumber
\\
Q_{aa}&=&\textit{T}+\chi,  Q_{bb}=\textit{T},
Q_{ab}=Q_{ba}=0,\nonumber
\end{eqnarray}
\end{subequations}
and
\begin{subequations}
\begin{eqnarray}
\lambda_{\pm}&=&\frac{1}{4}[\tilde{C}^{(\pm)}(-\Delta)+\tilde{C}^{(\pm)}(\Delta)],\nonumber
\\
\bar{\lambda}_{\pm}&=&\frac{1}{4}[\tilde{C}^{(\pm)}(-\Delta)-\tilde{C}^{(\pm)}(\Delta)]\nonumber.
\end{eqnarray}
\end{subequations}

\section{noise spectrum}

In this appendix, along the line of the conventional quantum jump
theory, we extend the method of noise spectrum calculation to the
unravelling approach developed in this work.
 Consider the correlation function  $E[dn(t+\tau)dn(t)]$,
 where
 $E[dn(t)]=\sum_{n_{1}}n_{1}{\rm Tr}[\mathcal{U}(n_{1},dt)\rho(t)]$.
 First, for the case $\tau>0$,

\begin{multline}
E[dn(t+\tau)dn(t)]
\\
\shoveleft{=\sum_{n_{1}}n_{1}{\rm
Prob}[dn(t)=n_{1}]E[dn_{c}(t+\tau)|_{dn(t)=n_{1}}],}
\end{multline}
where ${\rm Prob}[dn(t)=n_{1}]= {\rm
Tr}[\mathcal{U}(n_{1},dt)\rho(t)]$.
 At time $t+dt$ a definite value $n_{1}$ is picked out and the qubit
 state undergoes an immediate collapse,
 i.e. $\rho_{c}(t+dt)=\mathcal{U}(n_{1},dt)\rho(t)/{\rm Prob}[dn(t)=n_{1}]$.
During the time period $[t+dt,t+\tau]$, one can ignore the
measurement records owing to the ensemble nature of the
correlation function, therefore
\begin{multline}
E[dn_{c}(t+\tau)|_{dn(t)=n_{1}}]
\\
\shoveleft{=\sum_{n_{2}}n_{2} {\rm
Tr}[\mathcal{U}(n_{2},dt)e^{\mathcal{L}(\tau-dt)}
\frac{\mathcal{U}(n_{1},dt)\rho(t)}{{\rm Prob}[dn(t)=n_{1}]}].}
\end{multline}
To the leading order of $dt$ we have
\begin{widetext}
\begin{equation}
E[dn(t+\tau)dn(t)]=\sum_{n_{1}n_{2}}n_{2} {\rm
Tr}[\mathcal{U}(n_{2},dt)e^{\mathcal{L}\tau}
\times\frac{\mathcal{U}(n_{1},dt)\rho(t)}{{\rm
Tr}[\mathcal{U}(n_{1},dt)\rho(t)]}] n_{1}{\rm
Tr}[\mathcal{U}(n_{1},dt)\rho(t)] ={\rm
Tr}[\overline{\mathcal{U}}e^{\mathcal{L}\tau}\overline{\mathcal{U}}\rho(t)]dt^{2}
.
\end{equation}
\end{widetext}
Here we have introduced
$\overline{\mathcal{U}}\equiv\sum_{n}n\mathcal{U}(n,dt)/dt$. 
Next for $\tau=0$ we have
\begin{eqnarray}
E[dn(t)^{2}]&=&\sum_{n}n^{2}{\rm Prob}[dn(t)=n]\nonumber\\
&=&\sum_{n}n^{2}{\rm Tr}[\mathcal{U}(n,dt)\rho(t)]\nonumber
\\
&=&{\rm Tr}[\overline{\mathcal{U'}}\rho(t)]dt ,
\end{eqnarray}
where $\overline{\mathcal{U'}}=\sum_{n}n^{2}\mathcal{U}(n,dt)/dt$.
For short time $\tau$ this equal-time correlation will be
dominant, and $E[\frac{dn(t+\tau)}{dt}\frac{dn(t)}{dt}]$ can be
treated as $\delta$-correlated noise for a suitably defined
$\delta$ function. 
We thus obtain
\begin{eqnarray}
K_{I}(\tau)&=&E[\frac{dn(t+\tau)}{dt}\frac{dn(t)}{dt}]-E[\frac{dn(t+\tau)}{dt}]E[\frac{dn(t)}{dt}]
\nonumber
\\
&=&{\rm
Tr}[\overline{\mathcal{U}}e^{\mathcal{L}\tau}\overline{\mathcal{U}}\rho(t)]
-{\rm Tr}[\overline{\mathcal{U}}\rho(t+\tau)]{\rm
Tr}[\overline{\mathcal{U}}\rho(t)]\nonumber\\& &+{\rm
Tr}[\overline{\mathcal{U'}}\rho(t)]\delta(\tau).
\end{eqnarray}
Finally, it should be noted that $[dn(t)]^{2}\neq dn(t)$ in our
above treatment. This differs from the conventional quantum jump
theory where the stochastic number $dn(t)=0$ or 1 in the point
process.

\end{document}